\documentstyle[12pt]{article}

\begin{document}
\font\ticp=cmcsc10

\begin{titlepage}
\renewcommand{\thefootnote}{\fnsymbol{footnote}}

\rightline{DAMTP-R95/5}
\rightline{Phys. Rev. \ {D52.2} (1995).}
\vspace{0.1in}
\LARGE
\center{The String-Motivated Model }
\Large
\vspace{0.2in}
\center{}
\center{Justin D. Hayward\footnote{E-mail address:
J.D.Hayward@damtp.cam.ac.uk}}
\vspace{0.2in}
\large
\center{\em Department of Applied Mathematics and
Theoretical Physics,
\\ University of Cambridge,
\\  Silver Street, Cambridge CB3 9EW, U.K.}
\vspace{0.2in}

\today

\vspace{0.3in}

\begin{abstract}
The two-dimensional model which emerges from low-energy
 considerations of
 string theory
is written down.
 Solutions of this classical model are noted, including
 some examples which have
nontrivial tachyon field.
 One such represents the classical backreaction of the
tachyon field on the black
hole for a two parameter set of tachyon potentials.
 Assuming the classical black hole background in the
`Eddington-Finkelstein'
gauge, the tachyon equation is separable and the radial
part is solved by a
hypergeometric function, which is in general of complex
 argument.
 A semi-classical prescription for including the quantum
effects of the tachyon
field is described, and the resulting equations of
motion are found.
Special solutions of these equations are written down.

\vspace{0.3truecm}
\end{abstract}

\setcounter{footnote}{0}
\renewcommand{\thefootnote}{\arabic{footnote}}
\end{titlepage}
\section{Introduction}\setcounter{equation}{0}

String theory is thought to be important to
 the construction of quantum gravity.
The model that derives from string theory at tree level
in two dimensions
\cite{DEALIK} will be regarded here as a fundamental
field theory of gravity in its own right, and methods of
quantum field theory will be applied to it. This is
in contrast with taking the fundamental theory to be a
generally
reparametrisation invariant sigma model on
the two-dimensional world-sheet manifold of the string
and then
demanding that fields configure in such a way that is
consistent
 with conformal
invariance, {\it i.e.}, so that the $\beta$-functions
vanish.  To build up this theory
one would have to expand in world-sheet perturbation
theory, considering also
topologies, whereas
we shall work with the spacetime manifold. With this
distinction in
 procedures in mind, this model will be referred to as
the string-motivated model.

In semi-classical gravity the expectation value of
the  matter energy-momentum
tensor is coupled to the gravitational field.
If this coupling is to the Einstein tensor, then the
Bianchi Identities
 and energy
conservation ensure mathematical consistency\footnote
{In two dimensions, the result of applying this procedure to
the
Einstein Equation is either de Sitter or
 anti-de Sitter space. The backreaction
problem is thus completely soluble\cite{NS}.}.
 Physical consistency of this procedure
has been amusingly questioned in \cite{GEIK}.
Using this quantum principle of
equivalence, one has approximately included the effect of
the matter upon the geometry of the spacetime.
The aim is to see how a black hole
would develop when such back-reaction is considered.
 This would naturally extend
the original calculations\cite{SHA} in which the geometry of the spacetime
is treated as a fixed background. This has been done with some success
both generally\cite{SHM}, and in the context of several other dilaton gravity
models in two dimensions\cite{REVS}.

We begin the second section by introducing the string-motivated model,
and note classical solutions for
which the tachyon field is set to zero.
Two examples of solutions which have non-trivial tachyon field are then
found and written down. The first is flat space, the second represents
a naked singularity.
 Further examples of black holes which have undergone backreaction by
the tachyon field are given.
An ansatz for these black holes is applied, analogous
to the metric outside an evaporating star in
general relativity. The general solution is found in this form.
The ansatz shows the
position of the apparent horizon: its relationship to the
singularity and event horizon are calculable via a
certain integral.

 By ignoring backreaction
one can solve the field equation for the tachyon in the fixed
static black hole geometry. This has been done, for example in \cite{MK,ST},
 in the  Schwarzschild gauge assuming staticity.
 In the third section,  it is found that in the ingoing null
coordinates, for a particular tachyon potential, one obtains the same
hypergeometric equation for the radial part as in \cite{ST}, but there is
also a u-dependent piece.

In the fourth section, the procedure for coupling the
energy-momentum of the tachyon field to the field equations is
described. It is noted that this gives back the CGHS\cite{CGHS} equations
if one works in
the double-null coordinates, and drops tachyon terms.
 Thus the procedure used here is equivalent to adding
the Polyakov term for the tachyon field to the action itself.
 A set of semi-classical equations are found in the ingoing
`Eddington-Finkelstein'
gauge. Unfortunately, these equations are at least as complicated as
 those of \cite{CGHS},
where numerical methods were resorted to\cite{BGHS,JH3}, before the model was
adjusted so as to
be exactly soluble\cite{RST}.

\section{The String-Motivated Model}\setcounter{equation}{0}

The following is the action for the classical part of the string model
\begin{equation}
S = \frac {1} {2\pi} \int d^2 x \sqrt{-g} \,
 e^{-2\phi} \, ( R +
4{\nabla \phi}^2+4\lambda^2
- (\nabla T)^2 -V(T))
         -\frac 1 {\pi} \int d \Sigma \sqrt{-h} \,e^{-2\phi} \,
 (K-2{\nabla}_{n}\phi)  \label{eq:SMM}
\end{equation}

The fields present are the metric, $g_{\mu\nu}$, the dilaton,
$\phi$ and the tachyon $T$. There is a boundary term which makes
the variational problem well-defined and enables
 the thermodynamics of the theory to be derived.
 K is the trace of the second fundamental form
of the metric, and $\bf n$ is the normal vector to the boundary.
The equations of motion derived from (\ref{eq:SMM}) are
\begin{equation}
R_{\mu\nu} +2 \nabla_{\mu} \nabla_{\nu} \phi-
\nabla_{\mu} T \nabla_{\nu} T=0
\end{equation}

\begin{equation}
-R +4 (\nabla \phi)^2 -4 {\nabla}^2 \phi + {\nabla T}^2
+ V(T) - 4\lambda^2 =  0  \label{eq:DIL}
\end{equation}

\begin{equation}
{\nabla}^2 T - 2\nabla \phi \nabla T - \frac 1 2 {dV \over dT}=0
\label{eq:TAC}
\end{equation}
where $\lambda^2$ defines the mass scale here but is related to the
central charge in string theory, $V(T)$ is the tachyon potential.

Let us work in single null coordinates,
\begin{equation}
ds^2=-h(u,r) du^2-2dudr,  \label{eq:eBG}
\end{equation}
where h is a function on the spacetime to be determined.
 In these coordinates, the field equations are
\begin{equation}
h(h_{,rr}-2h_{,r}\phi_{,r})+2h_{,r}\phi_{,u}
-2h_{,u}\phi_{,r}+ 4 \phi_{,uu}-2 T_{,u}^{2}=0  \label{eq:BGUoo}
\end{equation}
\begin{equation}
h_{,rr}-2h_{,r}\phi_{,r} + 4 \phi_{,ur}-2 T_{,u}T_{,r}=0 \label{eq:BGUoi}
\end{equation}
\begin{equation}
2\phi_{,rr}- T_{,r}^{2}=0   \label{eq:BGUii}
\end{equation}

\begin{equation}
h_{,rr}-4h_{,r}\phi_{,r}-8\phi_{,r}\phi_{,u}+8 \phi_{,ur}
+4h{\phi_{,r}}^2-4 h\phi_{,rr}
+h {T_{,r}}^{2}-2 T_{,u}T_{,r}
=4{\lambda}^2-V         \label{eq:BDILU}
\end{equation}

\begin{equation}
h T_{,rr}-2T_{,ur}+h_{,r}T_{,r}-2h\phi_{,r}T_{,r}
+2\phi_{,u}T_{,r}+2\phi_{,r}T_{,u}-\frac 1 2 \frac {dV} {dT} =0
 \label{eq:eBT}
\end{equation}
\section{ Classical Solutions}\setcounter{equation}{0}
\subsection{$T=0$}

These equations simplify if one looks for solutions with
zero tachyon field.
There exists a timelike Killing vector in this case\cite{GIBPERB},
 and so it is
no  restriction to drop terms which contain derivatives of u.
One then has
\begin{equation}
h_{,rr}-2h_{,r}\phi_{,r}=0  \label{eq:BGUSTAToo}
\end{equation}

\begin{equation}
\phi_{,rr}=0  \label{eq:BGUSTATii}
\end{equation}

\begin{equation}
h_{,rr}-4h_{,r}\phi_{,r}
+4h{\phi_{,r}}^2-4 h\phi_{,rr}
-4{\lambda}^2=  0  \label{eq:BDILUSTAT}
\end{equation}
Thus there is a `linear dilaton' $\phi(r)=-\lambda r + \phi_0$
and a metric given by
\begin{equation}
h(r)=1-ae^{-2\lambda r},  \label{eq:hSTAT}
\end{equation}
where a is a constant.
The curvature information is in $R=-h_{,rr}=4a\lambda^2 e^{-2\lambda r}.$
There is a
curvature singularity at $r\to -\infty.$
It will be useful to transform solutions of the form (\ref{eq:eBG}) to null
coordinates. One transforms to conformally flat null coordinates via
\begin{equation}
ds^2=-\Omega^2(u,r) dudv=-h du^2-2dudr.  \label{eq:GLOB}
\end{equation}
If h is a function of r only, then the solution is $\Omega^2=h.$ A more
general
case is considered later.
One then finds that $h=(1\pm e^{-\lambda(v-u)})^{-1},$ where the positive sign
 corresponds
to $a>0$.
A further transform to `Kruskal' coordinates
\begin{equation}
\beta U=e^{-\lambda u}    \label{eq:KRi}
\end{equation}
\begin{equation}
\alpha V=e^{\lambda v}  \label{eq:KRii}
\end{equation}
yields, in the case of $a>0$, the familiar metric form for the
maximally-extended static black hole \cite{W},
\begin{equation}
ds^2=-\frac {dUdV} {-\frac {\lambda^2} {\alpha\beta} - \lambda^2 UV}.
 \label{eq:WI}
\end{equation}
where $\alpha\beta<0.$ If $\alpha\beta=-\frac {\lambda^3} {M}$ then M is
 the ADM mass\cite{CGHS}. If $a<0$ one finds
\begin{equation}
ds^2=-\frac {dUdV} { \frac {\lambda^2} {\alpha\beta} -{\lambda^2} UV }.
\label{eq:WII}
 \end{equation}
This latter solution represents a naked singularity, whereas
the former is the
black hole described in the first half of\cite{CGHS}.
The ground state solution, which has $M=0$, is the `linear dilaton vacuum'.
This corresponds to $a=0$ in the ingoing coordinate solution for h.

\subsection{\bf $T\neq 0$}

$\underline {Example 1}$

A first example is a flat geometry bathed in a
u-dependent tachyon.
One starts with a linear dilaton $\phi=-\lambda r$, and it is assumed that
$V(T)=aT^2.$
\begin{equation}
T=Ae^{\frac {-au} {2\lambda}} \label{eq:UTAC}
\end{equation}
\begin{equation}
h=1-\frac {V} {4\lambda ^2}   \label{eq:UTACi}
\end{equation}
Since h is a function of u only, this space is flat.
\vspace{10mm}

$\underline {Example  2}$

If $V=\lambda=0$, another set of solutions is of the form
\begin{equation}
h=\alpha r^{n}   \label{eq:POW}
\end{equation}
and the dilaton and tachyon are
\begin{equation}
\phi=-\frac 1 2  (1-n)\log r=\frac 1 2  T\sqrt {1-n}.  \label{eq:POWi}
\end{equation}
The curvature is $R=\a n(n-1)r^{(n-2)}.$
In null coordinates,
\begin{equation}
ds^2=-\alpha\Big(\frac {\alpha(1-n)} 2 \Big)^{\frac {n} {1-n}}
\frac {dudv} {(v-u)^{\frac {n} {n-1}}}.  \label{eq:POWii}
\end{equation}
In the case $n=2$ there is a singularity-free space of constant curvature.
For $n=0$ there is flat space and logarithmic $\phi=\frac 1 2  T. $
Otherwise, there is a timelike (naked) singularity on the line $v=u, r=0.$
\vspace{10mm}

$\underline {Example  3}$

In order to find solutions which represent the black hole perturbed
classically by the tachyon field,
one can assume that the metric takes the form of a black hole with a
dynamical horizon.
That is, solutions are of the form
\begin{equation}
ds^2=-h(u,r) du^2-2dudr,
\end{equation}
where
\begin{equation}
h=1-e^{-2\lambda(r-f(u))}  \label{eq:Ti}
\end{equation}

One then solves for the function f(u) which gives the
position of the horizon\footnote
{The motivation for this ansatz comes from four dimensional
 theory.
The metric outside
a radiating star is related to (\ref{eq:Ti}). This is called the Vaidya
metric,  and can be
written
\begin{equation}
ds^2=-(1-\frac {2M(u)} r)du^2-2dudr+r^2d\Omega ^2  \label{eq:VY}
\end{equation}
The mass in the Schwarzschild metric
has been upgraded from a constant to a function of the retarded time,
 which is reasonable if the radiation is made up
of massless particles.
The metric is a solution of the Einstein equations in a source field
of pure radiation,
\begin{equation}
G_{uu}=R_{uu}=8\pi T_{uu}=-\frac 2 {r^2} \frac {dM} {du}
\label{eq:eEIN}
\end{equation}
One might ask what is the future development of this system.
To answer this, consider Stefans Law,
\begin{equation}
\frac {dM} {du}=aAT^4\propto M^{-2}  \label{eq:eSTEF}
\end{equation}
where A is the area of the star, a is Stefans constant.
This implies that
\begin{equation}
\frac {dM} {du}\propto u^{-\frac 2 3}   \label{eq:eMU}
\end{equation}
The rate of mass decrease therefore diverges at finite retarded time.
This footnote will
be expanded upon elsewhere.}
$r_h=f(u).$
  The implicit assumption that
the horizon motion is u-dependent corresponds to the
masslessness of the
tachyon field.
 One might try to find solutions for which the dilaton
background is
linear, $\phi=-\lambda r$, but the field equations then
become
\begin{equation}
-2a \lambda^2 e^{2\lambda f} f_{,uu}  -e^{2\lambda r}T_{,u}^2=0  \label{eq:BUU}
\end{equation}
\begin{equation}
T_{,u}T_{,r}=0   \label{eq:BUR}
\end{equation}
\begin{equation}
T_{,r}^2=0,   \label{eq:BRR}
\end{equation}
which means $T=T(u)$. Inspection of (\ref{eq:BUU}) shows that $a=0$ so that
$T=0$, and no progress is made.

If, by contrast, one tries the following dilaton field:
\begin{equation}
\phi=-\lambda[r-f(u)],   \label{eq:DILi}
\end{equation}
then the value of the dilaton is thus fixed on the horizon.
The field equations become
\begin{equation}
2\lambda f_{,uu}- T_{,u}^2=0,   \label{eq:BGA}
\end{equation}
\begin{equation}
T_{,u}T_{,r}=0,     \label{eq:BGB}
\end{equation}
\begin{equation}
T_{,r}^2=0,   \label{eq:BGC}
\end{equation}
\begin{equation}
8\lambda^2 f_{,u} + V(T) = 0,  \label{eq:BDILA}
\end{equation}
\begin{equation}
4\lambda  T_{,u}= -{dV\over dT},   \label{eq:BTA}
\end{equation}

Equation (\ref{eq:BGB}) implies that T is a function of u only, and given the
 potential V(T), one can
solve for T in (\ref{eq:BTA}). One can substitute into equation
(\ref{eq:BGA}) and   integrate twice to obtain the function f and hence the
backreacted metric. \footnote{It should be noted that in this
form, the dilaton  field is
 constant on the
horizon. It can be shown however that the ADM mass of the solution is
related to
the dilaton there. The dilaton gives a coordinate
independent measure of position, so if it is constant on the horizon,
then the horizon
 is not moving. This suggests a static black hole, perturbed by the
classical tachyonic
backreaction. }

The tachyon potential is given by $V(T)= a T^2+ b T^3+...$
where $a$ and $b$ are taken from string theory calculations, and wont
 be specified here.

For  $V(T)=0$, equation (\ref{eq:BTA})  implies that T is a constant,
 and integrating up
(\ref{eq:BGA}) shows
that f(u) is then linear which is a static solution equivalent to
(\ref{eq:WI}),  the usual static black hole.

For quadratic V(T),
\begin{equation}
T= e^{\frac {-a} {2\lambda}(u+u_0)}.  \label{eq:QTi}
 \end{equation}
The solution for $f(u)$ is then
\begin{equation}
\lambda f(u)=\frac 1 {8\lambda} e^{\frac {-a} {\lambda}(u+u_0)} \label{eq:ROT}
\end{equation}
If the $O(T^3)$ term is included, one obtains
\begin{equation}
T=\frac {2a} {3b}  \frac 1 { e^{ \frac a {2\lambda} (u+u_0) }-1}.
\label{eq:ROTT}
\end{equation}
 Then
\begin{equation}
\lambda f(u) = A\left( \log |{ e^{ \frac a {2\lambda} (u+u_0) } -1}
 | + \frac { e^{ \frac a {2\lambda} (u+u_0) } }
{ (e^{ \frac a {2\lambda} (u+u_0) } -1)^2} \right)  \label{eq:ROTH}
\end{equation}
where A is a constant depending on a and b.

Note that these solutions (\ref{eq:QTi})-(\ref{eq:ROTH}) solve all the
field equations at once.

The geometry hasn't been fixed using the metric and dilaton equations in
isolation and setting $T=0$, as is done in the following section.

To see what these geometries look like globally, one can transform
to conformal null coordinates. Then the position of the horizon and
 singularity
are easily calculable. One must find $\Omega$ in
\begin{equation}
ds^2=-\Omega^2(u,r) dudv=-h du^2-2dudr.
\end{equation}
The following expression for $\Omega$ then obtains
\begin{equation}
4\Omega_{,u}=2h\Omega_{,r}-\Omega h_{,r}   \label{eq:PART}
\end{equation}
 For solutions of the form (\ref{eq:Ti}) one finds that
$\Omega^2=e^{-2\lambda(r+\frac u 2)}$ and
\begin{equation}
\lambda v= 2 e^{\lambda (r+\frac u 2)} - \lambda c(u)  \label{eq:SMV}
\end{equation}
where $c(u)=\int du e^{2\lambda f} e^{\lambda u}$.

Thus
\begin{equation}
 ds^2=-\frac {dudv} {{\frac \lambda 2}(c(u)+v)}   \label{eq:NULA}
\end{equation}
By rescaling $V=\lambda v$, and transforming to $U= -e^{-\lambda u}$,
one obtains the form
\begin{equation}
ds^2=\frac {dUdV} {\frac 1 2  U \lambda^2( V+\lambda c(U))}   \label{eq:KRUS}
\end{equation}

For $f=0$, the static Witten black hole given in equation ({\ref{eq:WI}) is
recovered.  If f is linear in u, this static metric still results.
Unfortunately, this transformation difficult to perform for the solutions
(\ref{eq:ROTH}) and (\ref{eq:ROT}),
except for the case $a=-\lambda^2$.
Since the form of the geometry is that of a black hole by ansatz, and the
 tachyon is a
scalar field so that it will remain non-trivial in any coordinate system:
 these are
black hole solutions with classical tachyon hair. The fact that the dilaton
 is constant at zero
on the horizon suggests that the solution is in fact static. The tachyon
field however is not constant on the horizon.

\section{ The Tachyon field in Fixed Geometry}
Now the static solution (\ref{eq:WI}) found by setting $T=0$ is fed into the
dilaton and gravitational field equations, and determine the tachyon equation.
 If one assumes that the solution is a separable function,
the radial part is found to be a hypergeometric function in r, as was
seen in \cite{ST},
 which reduces to an exponential function in flat space, while the
u-dependent piece is
exponential. If the constant of integration is taken to be imaginary,
this becomes
plane wave.
One can substitute the real solution back into the field equations,
 expanding
around the origin in r to try to find out how the tachyon backreacts
upon the geometry
perturbatively.

The tachyon equation of the string model was
\begin{equation}
{\nabla}^2 T - 2\nabla \phi \nabla T = \frac 1 2 {dV \over dT} \label{eq:BT}
\end{equation}
which in the coordinates (\ref{eq:GLOB}) with $f=0$  becomes
\begin{equation}
h T_{,rr} +2\lambda T_{,r}-\frac 1 2
\frac {dV} {dT} -2T_{,ru}-2\lambda T_{,u}=0
\label{eq:TU}
\end{equation}
Let $U=e^{\lambda r}T$, and look for solutions $U=\rho(r)\xi(u)$, assuming
 quadratic
tachyon potential with coefficient $a=-\lambda^2$.
One then finds that the function $\xi=e^{\frac c 2 u}$, where c is a constant.
 The equation for $ \rho(r)$ is
\begin{equation}
x(1-x)\rho ''+ (1+\frac {c} {2\lambda}-2x)\rho ' -\frac 1 4 \rho =0
\label{eq:HYPE}
\end{equation}
 This is a hypergeometric equation.
The solutions are
\begin{equation}
\rho=A F(\frac 1 2 ;\frac 1 2 ;1+\frac c {2\lambda};e^{-2\lambda r}) +
B F(\frac 1 2 ;\frac 1 2 ;1+\frac c {2\lambda};
1-e^{-2\lambda r})  \label{eq:HYSOLN}
\end{equation}
This gives T immediately.
We now return to the gravitational and dilaton equations.
 It is expected that the dilaton
and metric to be perturbed near the origin by this T field
which is fed into
the field equations. One finds that the new dilaton and
 metric must be static.
 The static field equations with $V(T)=aT^2$ are:

\begin{equation}
h_{,rr}-2h_{,r}\phi_{,r}=0
\end{equation}

\begin{equation}
2\phi_{,rr}- T_{,r}^{2}=0
\end{equation}

\begin{equation}
h_{,rr}-4h_{,r}\phi_{,r}
+4h{\phi_{,r}}^2-4 h\phi_{,rr}+h {T_{,r}}^{2}
-4{\lambda}^2+ a T^2=  0
\end{equation}

\begin{equation}
h T_{,rr}+h_{,r}T_{,r}-2h\phi_{,r}T_{,r}-aT =0  \label{eq:BTr}
\end{equation}
 T is known, hence (\ref{eq:BGUii}) implies $\phi(r)$ and (\ref{eq:BDILU})
acts  as a check for this
solution. One can calculate the power series
solution for metric and dilaton around the origin of r.
This naturally depends
on the expansion coefficients of the hypergeometrical tachyon,
 and is not very
instructive.

In summary, the black hole solution has been taken as a
fixed background in which
the tachyon moves. This gives a hypergeometric function.
When one iterates this
solution, one can find an expansion for the perturbed dilaton and metric
near the origin. The metric and dilaton must remain static,
 though in which global
configuration we do not know. The initial
assumption of $f(u)=0$ as a fixed background followed by many
iterations doesn't
necessarily lead to the result which would obtain if one were
 to solve the equations of motion at once, and is thus of limited value.

\section{Semi-Classical Treatment of the Model}

In this section the tachyon field is treated as a quantum field.
One simply adds to the expression for its classical stress
 tensor the quantum
stress tensor, which is derivable in two dimensions using
the trace
anomaly and the conservation equations. The
additional term might be produced by including a term in
the action. This
term is non-local, and it need not be specified here.
 The other fields are still treated classically, but one
would need later
to include dilaton and graviton loops. This question was
 addressed in the CGHS
model by proliferating the number of scalar fields, which
rendered other terms small and the semi-classical approximation exact.

The tachyon in (\ref{eq:SMM}) is not coupled as a standard scalar field,{\it
i.e.}
 \begin{equation}
{\cal L}=\sqrt {-g}((\nabla T)^2 - (m^2+\xi R)T^2)
\label{eq:LS}
\end{equation}
where $\xi$ is a numerical factor, which is zero for both minimal
and conformal couplings in two dimensions,
but rather,
\begin{equation}
{\cal L}_T=\sqrt {-g} e^{-2\phi}((\nabla T)^2+V(T))  \label{eq:LT}
\end{equation}

The lagrangian (\ref{eq:LT}) for the tachyon field is clearly conformally
coupled. The factor $e^{-2\phi}$ cannot be removed by a conformal
transformation in two dimensions.

Using dimensional analysis the trace anomaly for this
form of field must be
\begin{equation}
\alpha R + \beta  \label{eq:anom}
\end{equation}
where R is the Ricci scalar; $\alpha$ and $\beta$ are constants found in the
explicit  calculation through the heat equation.

By functionally differentiating the tachyon part of the Lagrangian with
respect to the metric one finds the classical stress tensor for the
tachyon is
\begin{equation}
T_{\mu\nu}=e^{-2\phi}(\nabla_{\mu} T\nabla_{\nu} T-\frac 1 2  g_{\mu\nu}
({\nabla T}^2 + V))  \label{eq:cIT}
\end{equation}
Using the field equations, this can be written
\begin{equation}
T_{\mu\nu}=2e^{-2\phi}(g_{\mu\nu}({\nabla \phi}^2-{\nabla}^2\phi-\lambda^2)
+\nabla_{\mu}\nabla_{\nu}\phi)  \label{eq:cITi}
\end{equation}
The simple step that we propose in order to include quantum
effects is to take the left hand side of this equation to be
the sum of the classical and quantum stress tensors for the
tachyon. For completeness, the equations of motion in the
gauge (\ref{eq:eBG})  are written down:-
\begin{equation}
e^{2\phi}T_{uu}=
hh_{,r}\phi_{,r}+h_{,r}\phi_{,u}-h_{,u}\phi_{,r}
-4h\phi_{,ru}+2h^2\phi_{,rr}-2h^2{\phi_{,r}}^2
+4h\phi_{,r}\phi_{,u}+2\phi_{,uu}+2h\lambda ^2   \label{eq:TI}
\end{equation}
\begin{equation}
e^{2\phi}T_{ur}=h_{,r}\phi_{,}-2\phi_{,ru}+2h\phi_{,rr}
-2h{\phi_{,r}}^2+4\phi_{,r}\phi_{,u}+2\lambda ^2  \label{eq:TII}
\end{equation}
\begin{equation}
e^{2\phi}T_{rr}=2\phi_{,rr}  \label{eq:TIII}
\end{equation}
where $T_{\mu\nu}=T^{cl}_{\mu\nu}+\langle {T^q_{\mu\nu}}\rangle.$

The classical components of $T_{\mu\nu}$ are
\begin{equation}
T^{cl}_{rr}=e^{-2\phi}T_{,r}^{2}  \label{eq:z}
\end{equation}
\begin{equation}
T^{cl}_{ur}=\frac 1 2  e^{-2\phi}(hT_{,r}^{2}+V(T))  \label{eq:y}
\end{equation}
\begin{equation}
T^{cl}_{uu}=e^{-2\phi}(T_{,u}^{2}+
\frac 1 2  h(hT_{,r}^{2}-2T_{,u}T_{,r}+V(T)))  \label{eq:x}
\end{equation}
If one works in the ingoing null coordinate gauge, it may be seen that it
isn't possible
to solve for the energy momentum tensor components for general
h(see (\ref{eq:eBG})), but
if one tries the ansatz (\ref{eq:Ti}),
the quantum piece of the energy momentum tensor can be found. If the
 trace anomaly
for the tachyon field is $\alpha R$, then this is
\begin{equation}
\langle {T^q_{rr}}\rangle=2\lambda ^2\alpha +\xi  \label{eq:TRR}
\end{equation}
\begin{equation}
\langle {T^q_{ur}}\rangle=-3\lambda ^2\alpha e^{-2\lambda(r-f)}
+\lambda ^2\alpha+
\frac 1 2 \xi(1-e^{-2\lambda(r-f)})  \label{eq:TUR}
\end{equation}
\begin{equation}
\langle {T^q_{uu}} \rangle=-e^{-2(r-f)}\Big(2\lambda^2\alpha \dot f+
4\lambda ^2\a+\frac 1 2 \xi\Big)
 +3\lambda^2\alpha e^{-4\lambda(r-f)}+\frac 1 4 \xi+t(u) \label{eq:TUU}
\end{equation}
where $\xi=Be^{2\lambda(2r+u)}$, and $t(u)$ is an arbitrary function
of u determined by
the boundary conditions. Keeping terms involving $\xi$,
there will be large distance
divergences in the components, so one sets $B=0$.

 These terms are added to the classical
stress tensor, and substituted into the field equations
(\ref{eq:TI})-(\ref{eq:TIII}).
 When $\alpha$ is set to zero,
one recovers the classical field equations (\ref{eq:BGUoo})-(\ref{eq:BDILU}).
These
 equations are clearly
quite complicated, and one cannot find closed form solutions.
Numerical solutions
might be interesting, but this is not pursued here.
\subsection{Solutions in the Conformal Gauge}
One can work in Kruskal double null coordinates,  i.e.(\ref{eq:GLOB}).
The equations of motion are then
\begin{equation}
e^{2\rho}(-4\lambda ^2 +V)-8\rho_{,uv}+
16\phi_{,uv}-16\phi_{,u}\phi_{,v}-4T_{,u}T_{,v}=0  \label{eq:BDILQ}
\end{equation}
\begin{equation}
\alpha e^{2\phi}(\rho_{,uu}-\rho_{,u}^2-t_{u}(u))+
4\rho_{,u}\phi_{,u}
-2\phi_{,uu}+T_{,u}^2=0   \label{eq:BQi}
\end{equation}
\begin{equation}
e^{2\rho}(-4\lambda ^2 +V)-4\alpha e^{2\phi}\rho_{,uv}+8\phi_{,uv}
-16\phi_{,u}\phi_{,v}=0  \label{eq:BQio}
\end{equation}
These equations reduce to the CGHS equations if one removes tachyon terms.

Another approach is to define
\begin{equation}
\Theta_{\mu\nu}^{cl}(\tilde T)=\Theta_{\mu\nu}^{cl}(T)+
\Theta_{\mu\nu}^{q}(T)    \label{eq:TT}
\end{equation}
where the quantity $\tilde T$ takes into account both the quantum
and classical
 contributions
to the energy-momentum tensor of the tachyon field. It is this
 field then that
appears in the
action (\ref{eq:SMM}). Taking $\tilde T=0$, so that $V(\tilde T)=0$,
one has the
classical CGHS equations with no matter. The solution to these is a one
 parameter family of
 static black holes, with
a vacuum state, the linear dilaton, given by the zero mass case
(\ref{eq:WI}).  But the relations
 (\ref{eq:TT}) will
give equations for the potential V(T) in terms of the conformal
 factor and the dilaton,
\begin{equation}
e^{-2\phi}V(T)=\alpha e^{-2\rho}\rho_{,uv}  \label{eq:VT}
\end{equation}
 which
are known, and which will determine the potential V(T) if one states
 the form of T.
 This will then
determine the constraint functions $t_{u}$ and $t_{v}$.
\begin{equation}
T_{,u}^2=\alpha e^{2\phi}(\rho_{,uu}-\rho_{,u}^2-t_{u}(u))
\label{eq:CONT}
 \end{equation}
and similarly for the advanced constraint equation in v.

 One could also choose the tachyon field to cancel out the
quantum piece after
 combining  (\ref{eq:BDILQ}) and (\ref{eq:BQio}), {\it i.e.}
\begin{equation}
T_{,u}T_{,v}=\alpha e^{2\phi}\rho_{,uv}  \label{eq:RSI}
\end{equation}
 Then if one works in the gauge $\rho=\phi$ and chooses
\begin{equation}
V=2\alpha\rho_{,uv}.  \label{eq:RSII}
\end{equation}
The
remaining equation is just the dynamical equation of the RST model.
\begin{equation}
-4\lambda^2e^{2\rho}-2\alpha e^{2\rho}\rho_{,uv}+8\rho_{,uv}-
16\rho_{,u}\rho_{,v}=0.  \label{eq:QS}
\end{equation}
These are the RST black holes, but generated by the tachyon field
and its potential.
The relations(\ref{eq:RSI}) and (\ref{eq:RSII}) imply the form of the
tachyon potential in terms of T.

To summarise, the equations for the
string-motivated model have been found,
which correspond to those of the CGHS model but in ingoing coordinates
 which give the
position of the apparent horizon. It seems that one has
to resort to
numerical solutions, where one could consider tachyonic ingoing matter,
 for various
potentials. Equilibrium static solutions and contact with `RST' black holes
are found by working in
 the conformal gauge and shaping the tachyonic terms.

\section{Conclusion}

One can try to simulate the black hole formation
 and evaporation in two dimensions: the hope is that the
results will have bearing upon more realistic descriptions,
as other scientific work in
two dimensions often has.

In this paper, a model of gravity arising from string theory is
treated as a quantum field theory.
First classical solutions are noted for zero and non-trivial tachyon field
configurations. Then, the equation of motion for the tachyon in a fixed flat
space and black hole geometry is solved, and is iterated into the
dilaton and gravitational field equations.
Finally, the quantum stress tensor for the tachyon field is found and
coupled appropriately to the classical field equations in another gauge
 from
that which has usually been used. The aim was to consider an analogous
coordinate system to that which highlights most clearly the behaviour
of a
radiating star in four dimensions. The solutions then immediately tell
us where the
apparent horizon is. Working in this gauge was useful in finding
classical solutions and considering the behaviour of the tachyon in a
 fixed
geometry. However, although this is another example of a coordinate
system in which one can calculate the quantum stress tensor components,
 and thus
derive semi-classical equations of motion, it does not yield simpler
equations than those found in the conformal gauge.

\section{Acknowledgements}

I thank Malcolm Perry for his ideas and criticisms, and Gary Gibbons
 for a later reading of the paper. I am grateful to the Cambridge Philosophical
Society for its financial support.

\end{document}